\documentclass[12pt]{article}
\usepackage{epsfig}
\usepackage{axodraw}
\newcommand{\mysection}{\setcounter{equation}{0}\section}
\renewcommand{\theequation}{\thesection.\arabic{equation}}

\def\beq{\begin{equation}}
\def\eeq{\end{equation}}
\def\beqa{\begin{eqnarray}}
\def\eeqa{\end{eqnarray}}

\begin{document}

\begin{center}
{\Large \bf Higher-order soft gluon corrections in single top quark \\
\bf production at the LHC}
\end{center}
\vspace{2mm}
\begin{center}
{\large Nikolaos Kidonakis}\\
\vspace{2mm}
{\it Kennesaw State University, Physics \#1202,\\
1000 Chastain Rd., Kennesaw, GA 30144-5591}\\
\end{center}

\begin{abstract}
I present a calculation of soft-gluon corrections to single  
top quark production in $pp$ collisions at the LHC via the  
Standard Model partonic processes in the $t$ and $s$ channels and 
associated top quark and $W$ boson production.
Higher-order soft-gluon corrections 
through next-to-next-to-next-to-leading order (NNNLO) are calculated at 
next-to-leading logarithmic (NLL) accuracy.
The soft-gluon corrections in the $s$ 
channel and in $tW$ production are large and dominant, 
while in the $t$ channel they are not a good approximation of the 
complete QCD corrections. 

\end{abstract}

\mysection{Introduction}

The production of top quarks in hadronic collisions can proceed via 
strong interaction processes
involving top-antitop pair production, or via 
electroweak processes involving 
the production of a single top (or antitop) quark. 
The top quark was first discovered via $t{\bar t}$ production at the 
Fermilab Tevatron by the CDF and D0 Collaborations in 1995 \cite{CDFtt,D0tt}. 
The observation of single top quark events has been more elusive but 
there has been recent evidence of such events by the D0 Collaboration 
\cite{D0st} with a cross section consistent with theoretical expectations 
\cite{NKsingletop}.
Single top quark production is an important production mode 
for further understanding of electroweak theory, measurement of the 
$V_{tb}$ CKM matrix element, and the search for new physics 
(see e.g. \cite{BBD,TT,TY,NKAB,WW}), so it is crucial to have accurate 
predictions for the cross sections. 

The Large Hadron Collider (LHC) at CERN has good potential for observation 
of single top quark events. 
The production of single top quarks in proton-proton collisions can 
proceed through three distinct partonic processes. 
The $t$-channel processes ($qb \rightarrow q' t$ 
and ${\bar q} b \rightarrow {\bar q}' t$, Fig. 1a) involve 
the exchange of a space-like $W$ boson, the $s$-channel 
processes ($q{\bar q}' \rightarrow {\bar b} t$, Fig. 1b) 
proceed via the exchange of a time-like $W$ boson,  
and associated $tW$ production ($bg \rightarrow tW^-$, Fig. 1c) 
involves the production of a top quark in association with a $W$ boson.
At the LHC the $t$-channel process is numerically the largest, 
the $s$-channel process is much smaller, and associated $tW$ production is 
intermediate in magnitude.

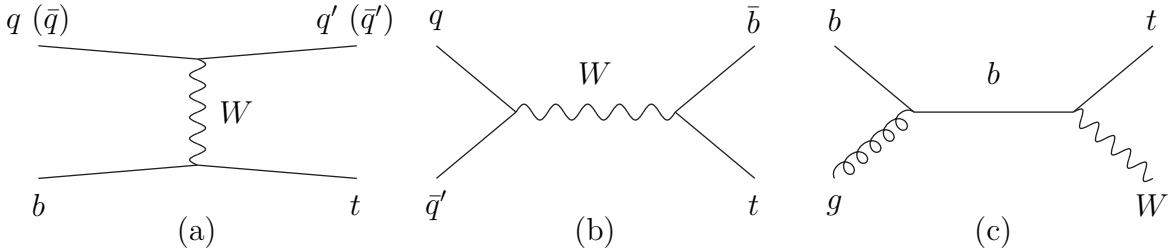
\begin{figure}[htb]
\begin{center}
\begin{picture}(120,120)(0,0)
\Line(0,75)(60,70)
\Line(60,70)(120,75)
\Line(0,25)(60,30)
\Line(60,30)(120,25)
\Photon(60,70)(60,30){3}{5}
\Text(0,15)[c]{$b$}\Text(0,85)[c]{$q$ (${\bar q}$)}
\Text(75,50)[c]{$W$}
\Text(120,15)[c]{$t$}\Text(120,85)[c]{$q'$ (${\bar q}'$)}
\Text(60,5)[c]{(a)}
\end{picture}
\hspace{8mm}
\begin{picture}(120,120)(0,0)
\Line(0,75)(30,50)
\Line(0,25)(30,50)
\Line(90,50)(120,75)
\Line(90,50)(120,25)
\Photon(30,50)(90,50){3}{5}
\Text(0,15)[c]{${\bar q'}$}\Text(0,85)[c]{$q$}
\Text(60,65)[c]{$W$}
\Text(120,15)[c]{$t$}\Text(120,85)[c]{${\bar b}$}
\Text(60,5)[c]{(b)}
\end{picture}
\hspace{8mm}
\begin{picture}(120,120)(0,0)
\Line(0,75)(30,50)
\Gluon(0,25)(30,50){3}{5}
\Line(30,50)(90,50)
\Photon(90,50)(120,25){3}{5}
\Line(90,50)(120,75)
\Text(0,15)[c]{$g$}\Text(0,85)[c]{$b$}
\Text(60,65)[c]{$b$}
\Text(120,15)[c]{$W$}\Text(120,85)[c]{$t$}
\Text(60,5)[c]{(c)}
\end{picture}
\end{center}
\vspace{-5mm}
\caption{\label{alo} Leading-order $t$-channel (a), $s$-channel (b), and  
associated $tW$ production (c) diagrams for 
single top quark production.}
\end{figure}

The next-to-leading order (NLO) QCD corrections to single top quark production 
have been calculated in Refs. \cite{BvE,SSW,Zhu,bwhl,zsull}.
In this paper we calculate the contribution of soft-gluon corrections 
to the cross section at the LHC beyond NLO. 
These corrections are normally dominant near partonic 
threshold for the production of a specified final state \cite{KS,LOS,NKtop} 
and they arise from incomplete cancellations of infrared 
divergences between virtual diagrams and real diagrams with soft 
(i.e. low-energy) gluons.
For the partonic process $f_1+f_2 \rightarrow t+X$ 
the kinematical invariants are 
$s=(p_1+p_2)^2$, $t=(p_1-p_t)^2$, $u=(p_2-p_t)^2$, 
$s_4=s+t+u-m_t^2-m_X^2$.
Near threshold, i.e. when we have just enough 
energy to produce the final state,  $s_4 \rightarrow 0$.
The threshold corrections then take the form of logarithmic plus 
distributions, $[\ln^l(s_4/m_t^2)/s_4]_+$, 
where $l\le 2n-1$ for the $n$-th order QCD corrections. 
These threshold corrections exponentiate; 
the resummation is carried out in moment space and it 
follows from the refactorization of the cross section into hard, soft, and 
jet functions that describe, respectively, the hard scattering, noncollinear 
soft gluon emission, and collinear gluon emission from the partons 
in the scattering \cite{KS,LOS}. 
To obtain a physical cross section, the moment-space result must be 
inverted to momentum space thus necessitating a prescription to 
handle the Landau singularity.
To avoid prescription ambiguities \cite{NKtop} we provide fixed-order 
expansions of the resummed cross section, as has been done for many other 
processes \cite{NKsingletop,NKAB,NKtop,KVtop,NKASV,NKNNNLO} 
(for a review see \cite{NKuni}). 

In the $n$-th order corrections in the strong coupling, $\alpha_s$, 
the leading logarithms (LL) are those 
with $l=2n-1$ while the next-to-leading logarithms (NLL) are those with 
$l=2n-2$. We calculate NLO, NNLO, and NNNLO soft-gluon threshold 
corrections at NLL accuracy, i.e. at each order including both leading and 
next-to-leading logarithms. 
We denote these corrections as NLO-NLL, NNLO-NLL, and NNNLO-NLL, respectively.
More details are given in Ref. \cite{NKsingletop}, where single top quark 
production at the Tevatron was discussed.
  
A calculation of threshold corrections is only meaningful if these 
corrections dominate the perturbative expansion, i.e. if they are a 
good approximation to the complete QCD corrections. 
This can be determined by comparing the NLO soft-gluon corrections 
with the complete NLO corrections.  
At Tevatron energies both top-antitop pair \cite{NKtop,KVtop} 
and single top quark \cite{NKsingletop} production satisfy this criterion. 
Since the LHC energy is much bigger it cannot be assumed that the same 
holds there. Moreover, since the kinematics and color flows are quite 
different for each channel, the validity of the threshold approximation 
must be examined for each channel seperately.
Indeed, a study of the corrections shows that while the threshold 
approximation works well for the $s$ channel and for $tW$ production, it 
is not valid for the $t$ channel at the LHC. 
Therefore for the $t$ channel we only update the NLO cross section.

In Section 2 we present numerical results for single top quark production via 
the $t$ channel at the LHC at NLO. Results including NNLO-NLL and NNNLO-NLL 
soft gluon corrections are provided for the 
$s$ channel in Section 3 and for associated $tW$ production in Section 4.
We note that in the $t$ and $s$ channels the cross section for single antitop 
production is different from that for single top production and we provide 
results for the antitop quark as well.
We use standard electroweak parameters \cite{PDG} throughout.

\mysection{Single top quark production via the $t$ channel at the LHC}

We begin with the $t$ channel. 
The dominant processes (with percentage contribution to the cross section) 
are $ub \rightarrow dt$ (68.1\% of the leading order cross section) and 
${\bar d} b \rightarrow {\bar u} t$ (11.5\%). Additional processes 
involving only quarks are $cb \rightarrow st$ (6.5\%) 
and Cabibbo-suppressed processes ($ub \rightarrow st$, 
$cb \rightarrow dt$, $us \rightarrow dt$, and further suppressed processes). 
Additional processes involving antiquarks and quarks 
are ${\bar s} b \rightarrow {\bar c}t$ (8.3\%) and Cabibbo-suppressed 
processes (${\bar d}b \rightarrow {\bar c}t$, 
${\bar s}b \rightarrow {\bar u}t$, ${\bar d} s \rightarrow {\bar u} t$, 
and further suppressed modes). 

A calculation of the NLO soft-gluon corrections shows that they are 
not a good approximation to the full NLO 
QCD corrections in the $t$ channel.  In fact they are large and negative 
while the exact NLO corrections \cite{bwhl} are small.  Therefore for the 
$t$ channel we only update the NLO results using the MRST 2004  
parton distribution functions (pdf) \cite{MRST2004} and we  
provide uncertainties for the cross section.

The NLO cross section for a top quark mass $m_t=175$ GeV is 
$\sigma^{t-{\rm channel}}_{\rm top} (m_t=175 \,{\rm GeV})=146 \pm 4 \pm 3$ pb, 
where the first uncertainty is from variation of the factorization
and renormalization scales, $\mu_F$ and $\mu_R$, between $m_t/2$ and 
$2m_t$, and the second uncertainty is from the pdf \cite{MRST2001E}. 
Adding these uncertainties in quadrature gives 
$\sigma^{t-{\rm channel}}_{\rm top} (m_t=175 \,{\rm GeV})=146 \pm 5$ pb.

The NLO cross section using the most recent value for the top quark 
mass, $m_t=171.4 \pm 2.1$ GeV \cite{topmass}, is  
$\sigma^{t-{\rm channel}}_{\rm top}(m_t=171.4 \pm 2.1 \,{\rm GeV})=150 \pm 5 
\pm 2.5 \, \pm 3$ pb, 
where the first uncertainty is due to the scale dependence, 
the second is due to the mass ($\pm 2.1$ GeV), and the third is due to the 
pdf. Adding these uncertainties in quadrature we find that 
$\sigma^{t-{\rm channel}}_{\rm top}(m_t=171.4 \pm 2.1 \,{\rm GeV})=150 \pm 6$ 
pb.  

The cross section for the production of an antitop quark 
at the LHC is different from that for a top quark.
The corresponding results for the two choices of mass are 
$\sigma^{t-{\rm channel}}_{\rm antitop} (m_t=175 \,{\rm GeV})=89 \pm 3 \pm 2$ 
pb $=89 \pm 4$ pb and 
$\sigma^{t-{\rm channel}}_{\rm antitop} (m_t=171.4 \pm 2.1 \,{\rm GeV})=92 
\pm 3 \pm 1.5 \pm 2$ pb $=92 \pm 4$ pb.

\mysection{Single top quark production via the $s$ channel at the LHC}

We continue with the $s$ channel.
The dominant process is $u {\bar d} \rightarrow {\bar b} t$ (85.0\% of the 
leading-order cross section).
Additional processes are $c {\bar s} \rightarrow {\bar b} t$ (10.2\%) 
and Cabibbo-suppressed processes ($u {\bar s} \rightarrow {\bar b} t$, 
$c{\bar d} \rightarrow {\bar b}t$, $u{\bar d} \rightarrow {\bar s}t$,  
and further suppressed modes). 

\begin{table}[htb]
\begin{center}
\begin{tabular}{|c|c|c|c|c|} \hline
$s$ channel top & LO & NLO approx & NNLO approx & NNNLO approx \\ \hline
$m_t=170$  & 5.40 & 7.53  & 8.08  & 8.33 \\ \hline
$m_t=172$  & 5.18 & 7.23  & 7.76  & 8.00 \\ \hline
$m_t=175$  & 4.87 & 6.79  & 7.29  & 7.52 \\ \hline
\end{tabular}
\caption[]{The leading-order and approximate higher-order 
cross sections for top quark production in the $s$ channel in pb 
for $pp$ collisions with $\sqrt{S} = 14$ TeV and 
$m_t=170$, 172, and 175 GeV. We use the MRST2004 NNLO pdf \cite{MRST2004}
and we set $\mu_F=\mu_R=m_t$.}
\end{center}
\end{table}

In Table 1 we give results for the LO cross section and for 
the approximate NLO, NNLO, and NNNLO cross sections which include 
the threshold soft-gluon corrections at NLL accuracy.
The soft-gluon corrections are relatively large for this channel; 
this is also true for the exact NLO corrections \cite{bwhl}. 
We also note that the approximate NLO cross section is only about 4\% 
larger than the exact NLO result, showing that 
the threshold corrections are dominant and provide the bulk of the QCD 
corrections and, thus, that the threshold approximation works quite well.
 
As shown in Table 1, the NNNLO approximate cross section 
for $m_t=175$ GeV is 7.52 pb. 
After matching to the exact NLO cross section \cite{bwhl}, 
we find that the matched 
cross section (i.e. exact NLO plus NNLO-NLL and NNNLO-NLL 
threshold corrections) is 
$\sigma^{s-{\rm channel}}_{\rm top}(m_t=175 \,{\rm GeV})
=7.23^{+0.53}_{-0.45}\pm 0.13$ pb, where the first uncertainty is due to 
the scale dependence and the second is due to the pdf.
Adding these uncertainties in quadrature we find
 $\sigma^{s-{\rm channel}}_{\rm top}(m_t=175 \,{\rm GeV})
=7.23^{+0.55}_{-0.47}$ pb.

The NNNLO approximate cross section for $m_t=171.4$ GeV is 8.10 pb. 
If we match this to the exact NLO cross section, then we find 
$\sigma^{s-{\rm channel}}_{\rm top}(m_t=171.4 \pm 2.1 \, {\rm GeV})=
7.80^{+0.58}_{-0.48} {}^{+0.36}_{-0.33} \pm 0.14 \; {\rm pb}$, 
where the first uncertainty is due to scale dependence, the second is due to 
the mass ($\pm 2.1$ GeV), and the third is the pdf uncertainty.
Adding these uncertainties in quadrature we find that 
$\sigma^{s-{\rm channel}}_{\rm top}(m_t=171.4 \pm 2.1\, {\rm GeV})
=7.80^{+0.70}_{-0.60}$ pb.

\begin{figure}
\begin{center}
\hspace{-5mm}
\includegraphics[width=8cm]{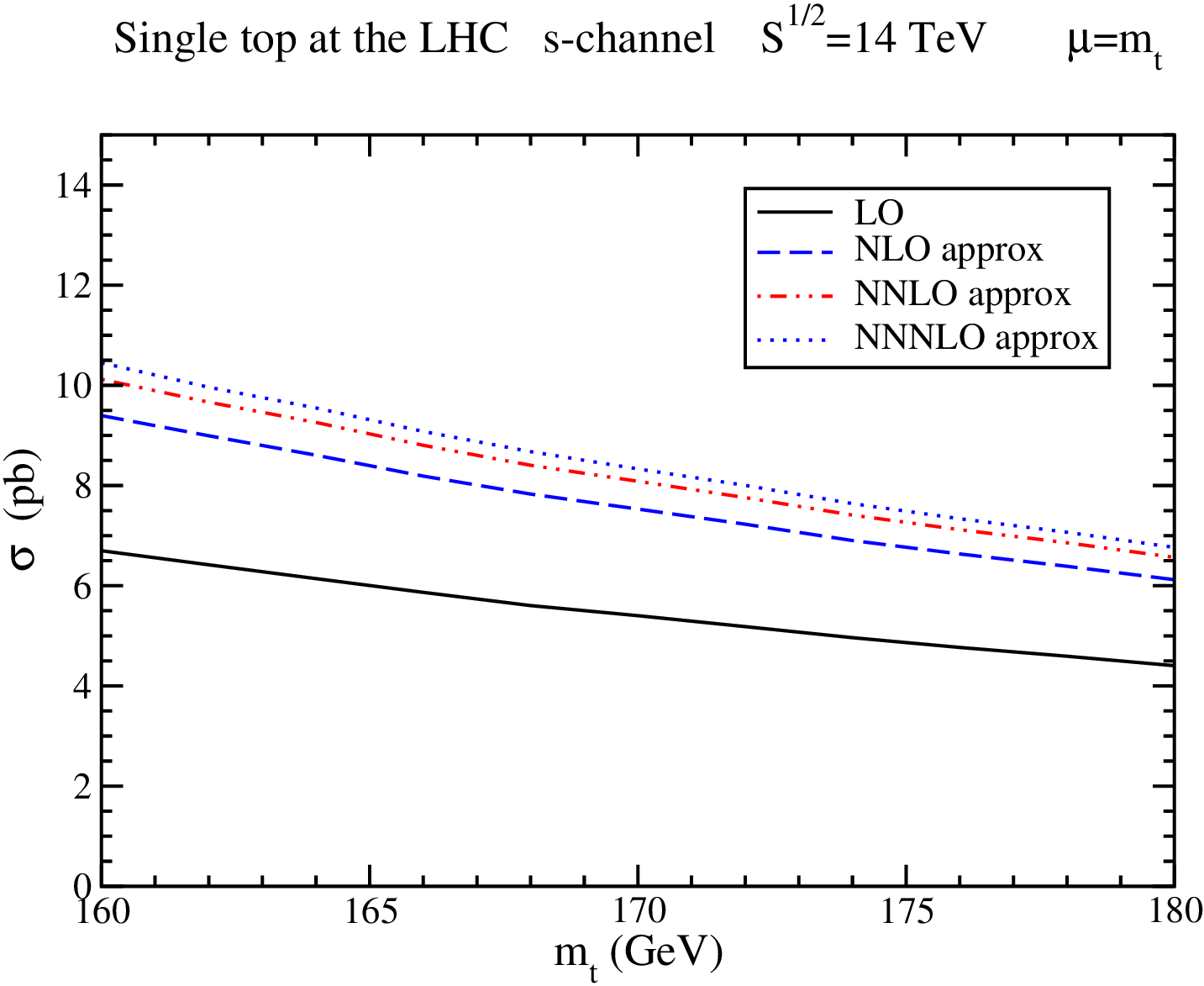}
\hspace{5mm}
\includegraphics[width=8cm]{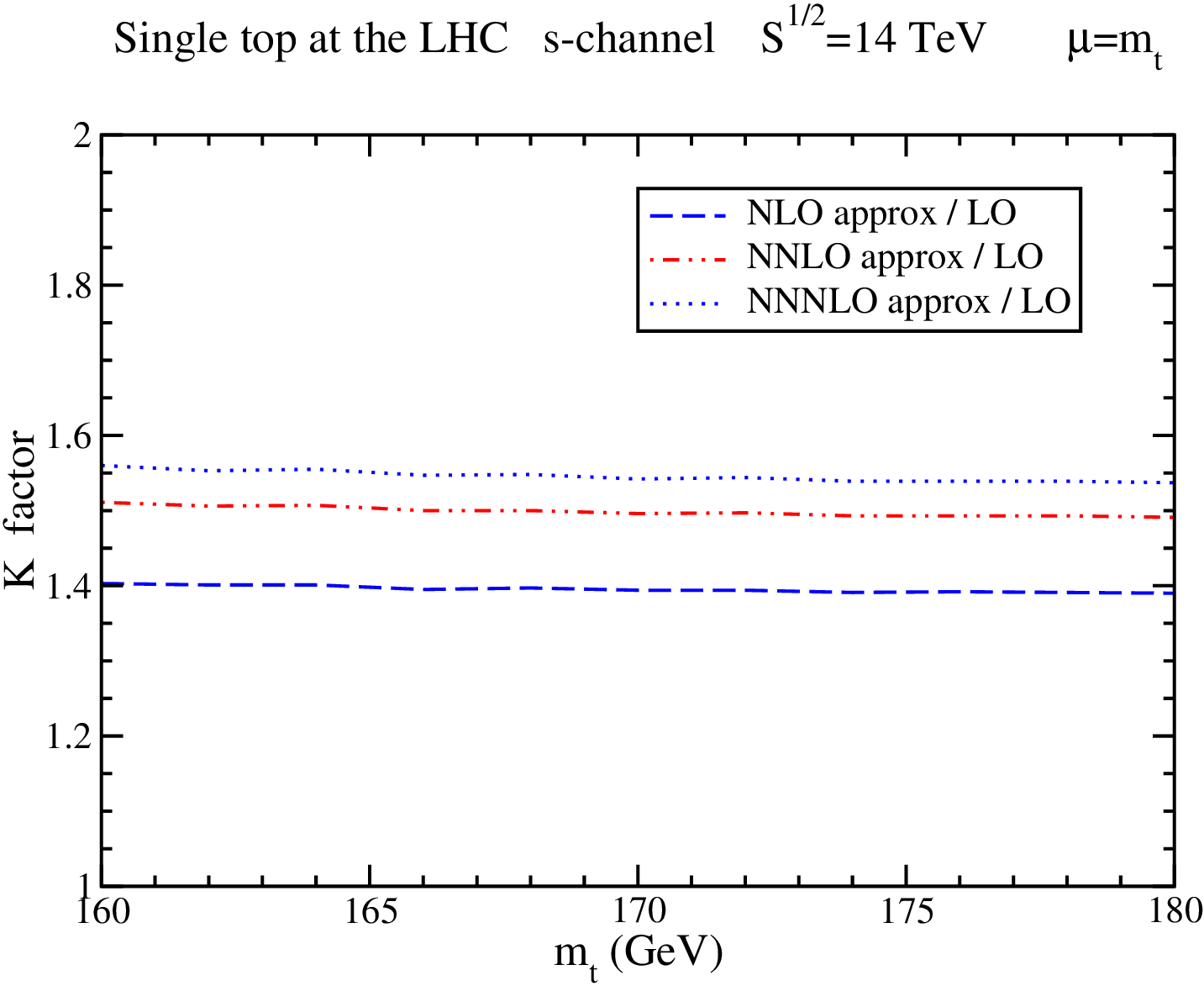}
\caption{Left: The cross section for single top production at the LHC  
in the $s$ channel. Here $\mu=\mu_F=\mu_R=m_t$.
Right: the $K$ factors.}
\label{schlhcmtplot}
\end{center}
\end{figure}

In Fig. \ref{schlhcmtplot} we plot the cross section 
for single top quark production in the $s$ channel 
at the LHC with $\sqrt{S}=14$ TeV  setting 
both the factorization and renormalization scales to $\mu=m_t$. 
On the left-hand side we plot the LO and approximate higher-order 
cross sections as functions of the top quark mass. 
On the right-hand side we show the $K$ factors, defined as the ratios 
of the higher-order cross sections to the LO cross section.
The $K$ factors are quite large, thus showing that the corrections provide a 
big enhancement to the cross section, and are fairly 
constant over the top-quark mass range shown. As seen in the plot, 
the NLO-NLL corrections provide a 40\% increase of the LO cross section, the 
NNLO-NLL corrections provide an additional 10\%, and the NNNLO-NLL corrections 
a further 5\%.

\begin{table}[htb]
\begin{center}
\begin{tabular}{|c|c|c|c|c|} \hline
$s$ channel antitop & LO & NLO approx & NNLO approx & NNNLO approx \\ \hline
$m_t=170$  & 3.44 & 4.22 & 4.14 & 4.09  \\ \hline
$m_t=172$  & 3.30 & 4.05 & 3.96 & 3.92 \\ \hline
$m_t=175$  & 3.08 & 3.78 & 3.71 & 3.67 \\ \hline
\end{tabular}
\caption[]{The leading-order and approximate higher-order 
cross sections for antitop quark production in the $s$ channel in pb 
for $pp$ collisions with $\sqrt{S} = 14$ TeV and 
$m_t=170$, 172, and 175 GeV. We use the MRST2004 NNLO pdf 
\cite{MRST2004} and we set 
$\mu_F=\mu_R=m_t$.}
\end{center}
\end{table}

The cross section for single antitop quark production in the $s$ channel 
at the LHC is different from that for single top quark production. 
In Table 2 we give results for antitop 
quark production at the LHC.
The approximate NLO cross section is only 
about 10\% smaller than the exact NLO result \cite{bwhl}, 
showing again that the threshold corrections provide the 
bulk of the QCD corrections.
The soft-gluon corrections beyond NLO are small and negative.

As shown in Table 2, the NNNLO approximate cross section for 
$m_t=175$ GeV is 3.67 pb. After matching to the exact NLO cross 
section \cite{bwhl} we find
$\sigma^{s-{\rm channel}}_{\rm antitop}(m_t=175 \,{\rm GeV})
=4.03^{+0.10}_{-0.12}\pm 0.10$ pb, where the first uncertainty is due 
to the scale dependence and the second is due to the pdf.
Adding these uncertainties in quadrature we find 
$\sigma^{s-{\rm channel}}_{\rm antitop}(m_t=175 \,{\rm GeV})
=4.03^{+0.14}_{-0.16}$ pb.

The NNNLO approximate cross section for $m_t= 171.4$ GeV is 3.96 pb. 
If we match this to the exact NLO cross section, we find
$\sigma^{s-{\rm channel}}_{\rm antitop}(m_t=171.4 \pm 2.1 \, {\rm GeV})=
4.35^{+0.11}_{-0.13} {}^{+0.21}_{-0.19} \pm 0.11 \; {\rm pb}$, 
where the first uncertainty is due to scale dependence, the second is 
due to the mass ($\pm 2.1$ GeV), and the third is the 
pdf uncertainty.
Adding these uncertainties in quadrature we find that 
$\sigma^{s-{\rm channel}}_{\rm antitop}(m_t=171.4 \pm 2.1\, {\rm GeV})
=4.35 \pm 0.26$ pb.

\begin{figure}
\begin{center}
\hspace{-5mm}
\includegraphics[width=8cm]{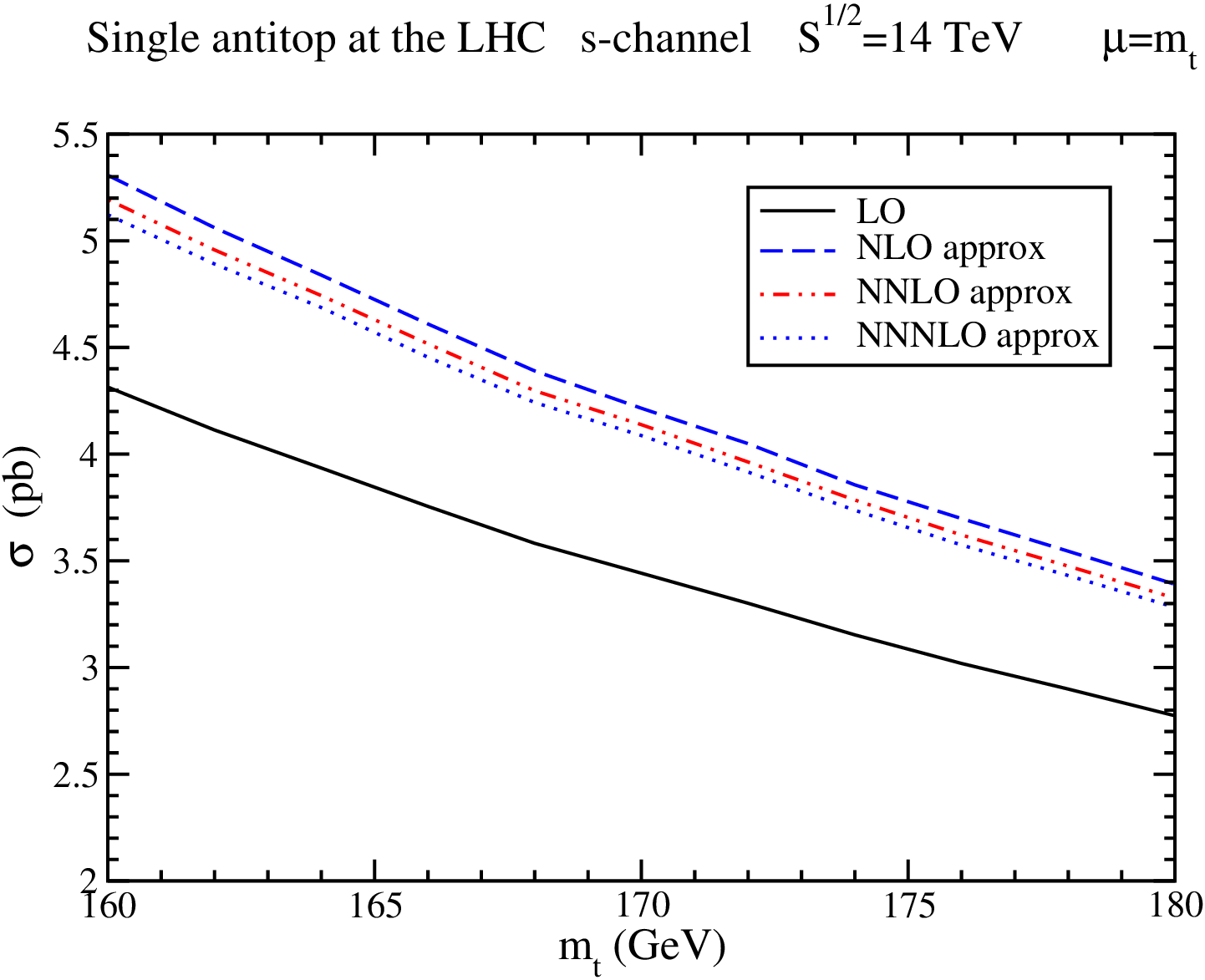}
\hspace{5mm}
\includegraphics[width=8cm]{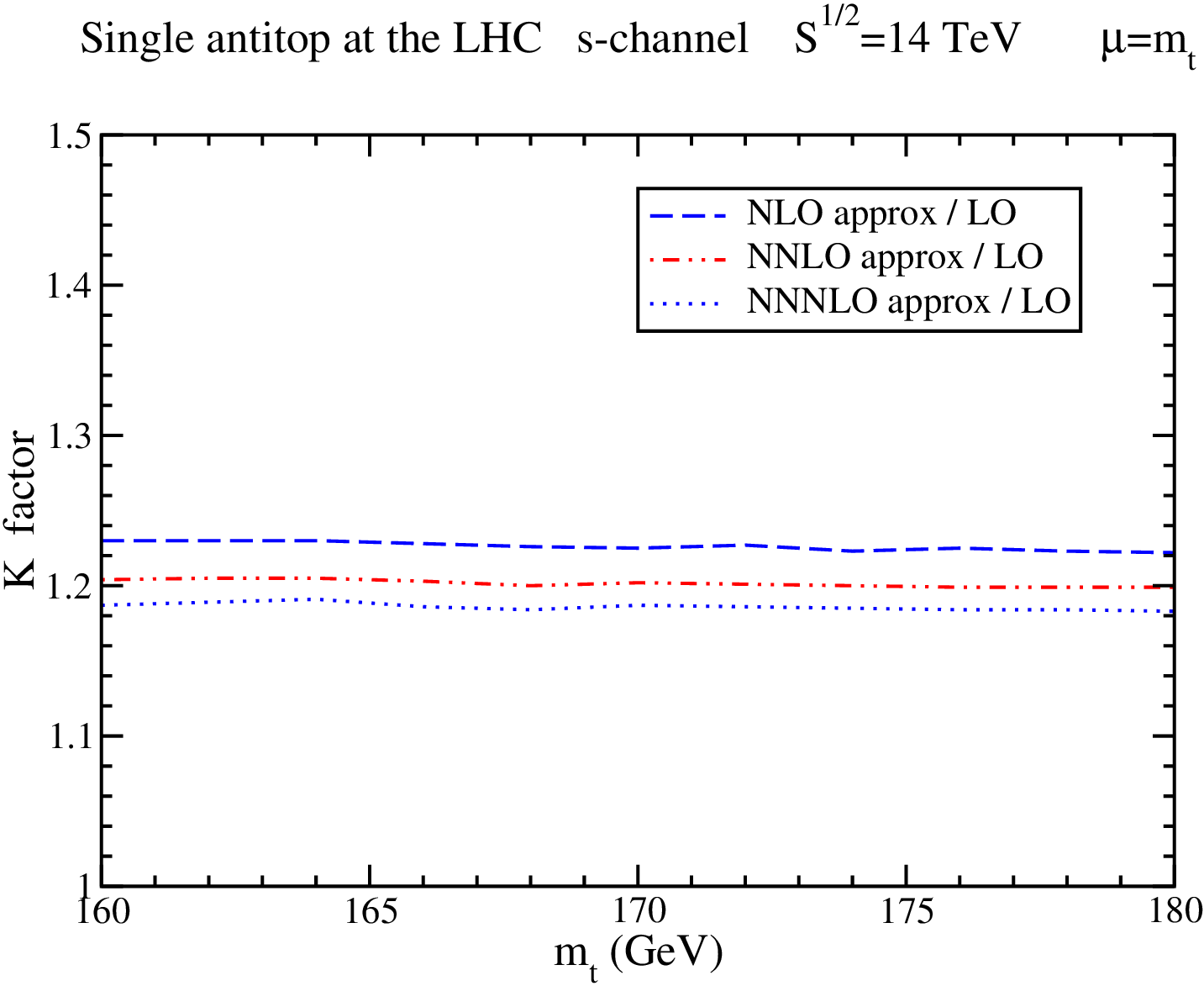}
\caption{Left: The cross section for single antitop production at the LHC  
in the $s$ channel. Here $\mu=\mu_F=\mu_R=m_t$.
Right: The $K$ factors.}
\label{aschlhcmtplot}
\end{center}
\end{figure}

In Fig. \ref{aschlhcmtplot} we plot the cross section 
for single antitop quark production in the $s$ channel 
at the LHC with $\sqrt{S}=14$ TeV setting 
both the factorization and renormalization scales to $\mu=m_t$. 
On the left-hand side we plot the LO and approximate 
higher-order cross sections versus the top quark mass.
The $K$ factors are shown on the right. 
They are not as big as for single top production in this channel and 
in fact the NNLO-NLL and NNNLO-NLL corrections are small and negative.
As seen in the plot, the NLO corrections provide a 23\% increase of the 
LO cross section, the NNLO corrections provide an additional -3\%, and 
the NNNLO corrections a further -1.5\%.

\mysection{Associated $tW$ production at the LHC}

Associated $tW$ production proceeds via $bg \rightarrow tW^-$ 
(Cabibbo-suppressed contributions from 
$sg\rightarrow tW^-$ and $dg\rightarrow tW^-$ are negligible). 

\begin{table}[htb]
\begin{center}
\begin{tabular}{|c|c|c|c|c|} \hline
$tW$ production & LO & NLO approx & NNLO approx & NNNLO approx \\ \hline
$m_t=170$  & 32.1 & 44.3 & 48.6 & 51.0 \\ \hline
$m_t=172$  & 31.2 & 43.1 & 47.4 & 49.6 \\ \hline
$m_t=175$  & 29.9 & 41.3 & 45.2 & 47.3 \\ \hline
\end{tabular}
\caption[]{The leading-order and approximate higher-order 
cross sections for associated $tW$ production in pb 
for $pp$ collisions with $\sqrt{S} = 14$ TeV and 
$m_t=170$, 172, and 175 GeV. We use the MRST2004 NNLO pdf \cite{MRST2004}
and we set $\mu_F=\mu_R=m_t$.}
\end{center}
\end{table}

In Table 3 we give results for the LO and the higher-order cross sections 
including the threshold soft-gluon corrections at NLL accuracy at each order.
The soft-gluon corrections are relatively large for this channel, even  
more than in the $s$ channel, and they provide the bulk of the QCD corrections.
 
As shown in Table 3, the NNNLO approximate cross section for 
$m_t=175$ GeV is 47.3
pb. After matching to the exact NLO cross section \cite{Zhu} we find
$\sigma^{tW}(m_t=175 \,{\rm GeV})
=41.1 \pm 4.1 \pm 1.0$ pb, where the first uncertainty is due 
to the scale dependence and the second is due to the pdf.
Adding these uncertainties in quadrature we find 
$\sigma^{tW}(m_t=175 \,{\rm GeV})=41.1 \pm 4.2$ pb.

The NNNLO approximate cross section for $m_t=171.4$ GeV is 50.1 pb. 
If we match this to the exact NLO cross section then we find
$\sigma^{tW}(m_t=171.4 \pm 2.1 \, {\rm GeV})=43.5 \pm 4.5 \pm 1.5 \pm 1.0$ pb,
where the first uncertainty is due to scale dependence,  
the second is due to the mass ($\pm 2.1$ GeV), and the third is the 
pdf uncertainty.
Adding these uncertainties in quadrature we find that 
$\sigma^{tW}(m_t=171.4 \pm 2.1 \, {\rm GeV})=43.5 \pm 4.8$ pb.

\begin{figure}
\begin{center}
\hspace{-5mm}
\includegraphics[width=8cm]{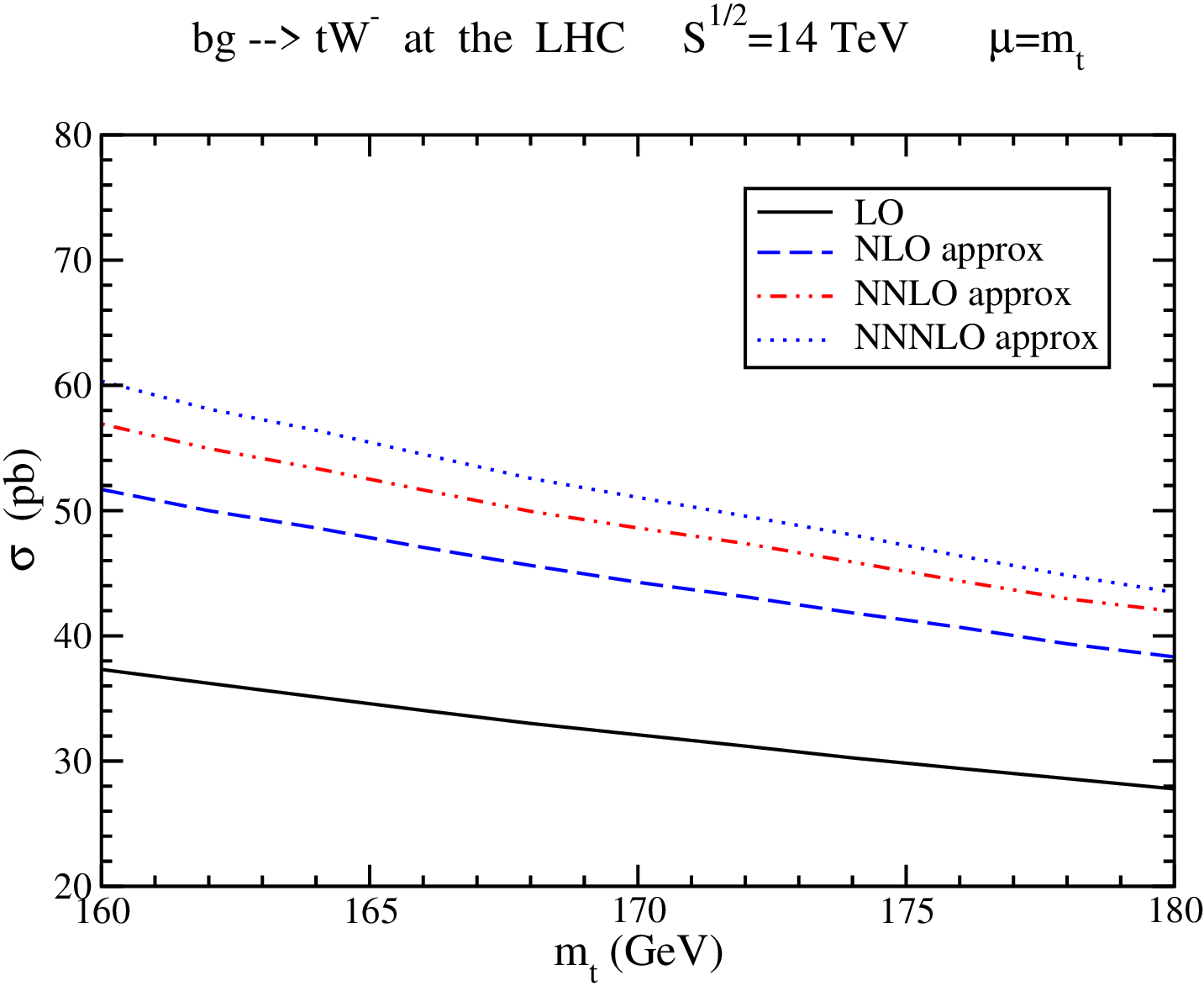}
\hspace{5mm}
\includegraphics[width=8cm]{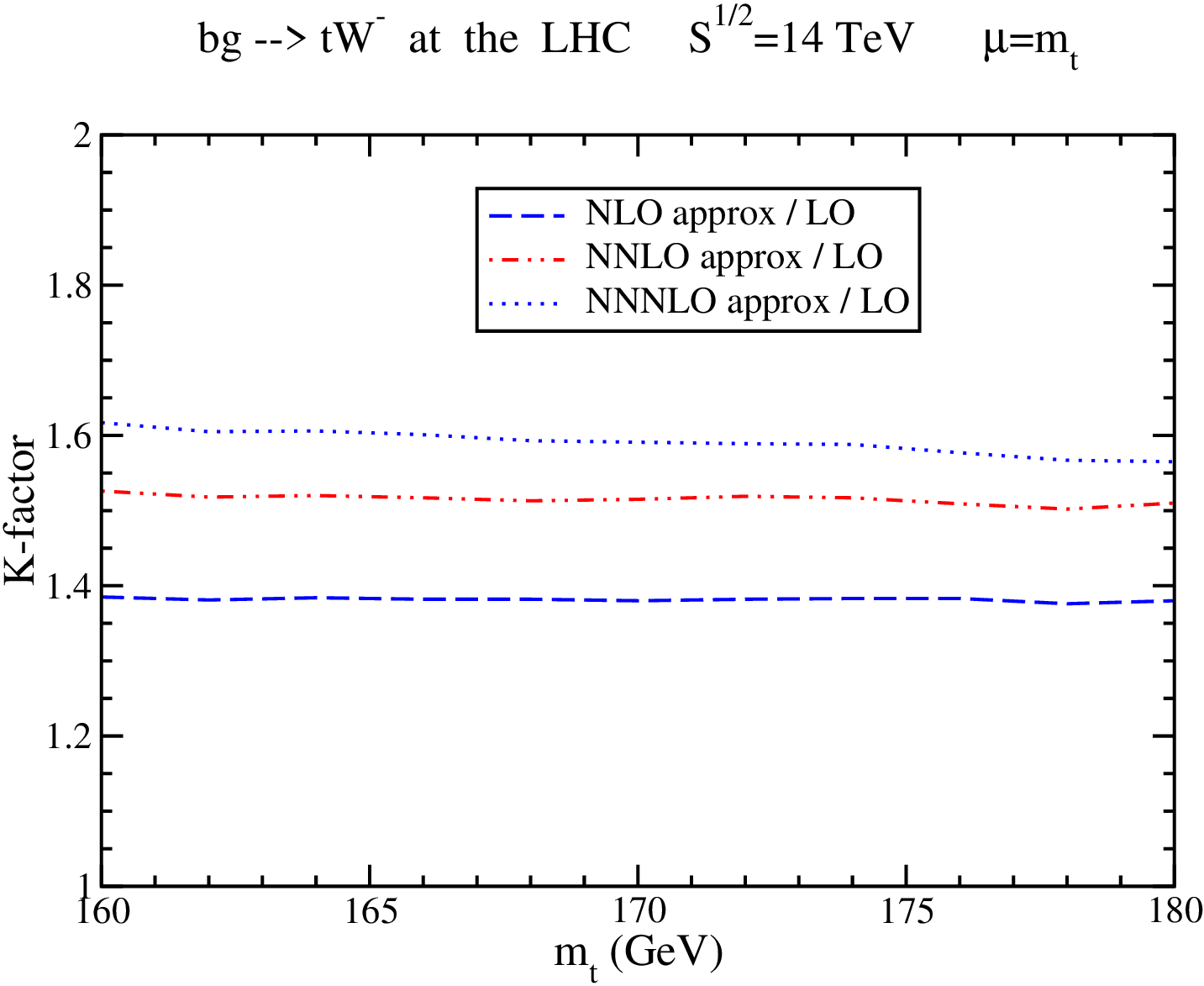}
\caption{Left: The cross section for associated $tW$ production at the 
LHC. Here $\mu=\mu_F=\mu_R=m_t$. Right: The $K$ factors.}
\label{bglhcmtplot}
\end{center}
\end{figure}

In Fig. \ref{bglhcmtplot} we plot the cross section 
for associated $tW$ production at the LHC with $\sqrt{S}=14$ TeV 
setting the scales to $\mu=m_t$. 
On the left-hand side we plot the LO and approximate higher-order 
cross sections, and on the right-hand side we plot the $K$ factors. 
The latter are quite large, showing that the soft-gluon corrections 
provide a big enhancement to the cross section, and fairly 
constant over the top-quark mass range shown. As seen from the plot, 
the NLO corrections provide a 38\% increase of the LO cross section, the 
NNLO corrections provide an additional 14\%, and the NNNLO corrections 
a further 7\%.

Finally, we note that the cross section for the associated production of an 
antitop quark is identical to that for a top quark.

\mysection{Conclusion}

We have studied single top quark production at the LHC and 
have calculated the soft-gluon
corrections to the cross section in three different production channels.
The results differ a lot among channels. In the $t$ channel the 
soft-gluon corrections are not a good approximation of the full NLO result.
The cross section for $t$-channel single top quark production at the LHC is  
$\sigma^{t-{\rm channel}}_{\rm top}(m_t=171.4 \pm 2.1 \,{\rm GeV})
=150 \pm 6$ pb, 
where the uncertainty indicated includes the scale dependence, 
the uncertainty in the top quark mass, and the pdf uncertainty.
For single antitop quark production the cross section is 
$\sigma^{t-{\rm channel}}_{\rm antitop}(m_t=171.4 \pm 2.1 \,{\rm GeV})
=92 \pm 4$ pb.

In the $s$ channel, however, the NLO soft-gluon corrections are a good 
approximation of the full NLO result and they are significant.  
The threshold approximation works well and the higher-order 
NNLO and NNNLO soft-gluon contributions, at NLL accuracy, 
provide further enhancements.
Our best estimate for the single top quark cross section in this channel 
at the LHC is 
$\sigma^{s-{\rm channel}}_{\rm top}(m_t=171.4 \pm 2.1\, {\rm GeV})
=7.80^{+0.70}_{-0.60}$ pb. For the single antitop quark the cross section is
$\sigma^{s-{\rm channel}}_{\rm antitop}(m_t=171.4 \pm 2.1 \, {\rm GeV})
=4.35 \pm 0.26$ pb.

The threshold soft-gluon corrections for associated $tW$ production are 
also large.
Our best estimate for the single top quark cross section in this channel 
at the LHC is 
$\sigma^{tW}(m_t=171.4 \pm 2.1 \, {\rm GeV})=43.5 \pm 4.8$ pb.
The cross section is the same for associated antitop production.

\mysection*{Acknowledgements}

This work has been supported by the National Science Foundation under 
grant PHY 0555372.

\setcounter{equation}{0}
\renewcommand{\theequation}{A.\arabic{equation}}

\end{document}